\documentclass[prc,preprint,floatfix]{revtex4}

\usepackage{amsmath}
\usepackage{graphicx}

\newcommand{\be}{\begin{eqnarray}}
\newcommand{\ee}{\end{eqnarray}}

\begin{document}

\title{The Effect of Shear Viscosity on Spectra, Elliptic Flow, and HBT Radii} 

\author{Derek Teaney}
\email{dteaney@quark.phy.bnl.gov}
\affiliation{%
 Physics Department, Brookhaven National Laboratory,
                Upton, N.Y. 11973, U.S.A.
}

\date{\today}

\begin{abstract}
I calculate the first  correction  to the
thermal distribution function of an expanding gas due 
to shear viscosity.  With this
modified distribution function I estimate viscous corrections to
spectra, elliptic flow, and HBT radii in hydrodynamic simulations of
heavy ion collisions using the 
blast wave model. For reasonable values of the shear viscosity,
viscous corrections become of order one when the transverse momentum of
the particle is larger than 1.7 GeV. 
This places a bound on the $p_{T}$
range accessible to hydrodynamics for this observable. Shear corrections
to elliptic flow cause $v_{2}(p_{T})$ to veer below the ideal 
results for $p_{T} \approx 0.9$ GeV.  Shear corrections to the
longitudinal HBT radius
$R^{2}_{L}$ are large and negative. 
The reduction of $R_{L}^2$
can be traced to the reduction of the longitudinal pressure. 
Viscous corrections cause the longitudinal radius to deviate
from the $\frac{1}{\sqrt{m_T}}$ scaling which is observed in the data 
and which is predicted by ideal hydrodynamics.
The correction to the sideward radius $R^{2}_{S}$ is small. 
The correction to the outward radius $R^{2}_{O}$
is also negative and tends to make $R_{O}/R_{S} \approx 1$. 
\end{abstract}

\maketitle

\section{Introduction}
One of the most exciting results of the Relativistic Heavy Ion Collider
(RHIC) is the observation of collective motion. 
In particular, the experiments have
measured
a large elliptic flow in non-central collisions 
\cite{ v2-star3, v2-star2, v2-star1,v2-phenix,v2-phobos}. 
Elliptic flow is quantified with
the second harmonic of the azimuthal distribution of produced particles
\begin{equation}
v_2(p_{T}) =  \langle \cos(2\phi) \rangle_{p_T}
    \equiv 
    \frac{
     \int^{\pi}_{-\pi} d\phi \,\cos(2\phi)  \,
                \frac{d^3N }{ dy\, p_t\,dp_t\,d\phi}
        }{
       \int^{\pi}_{-\pi}\, d\phi\,  
                \frac{d^3N }{ dy\,p_t\,dp_t\,d\phi}  
        }\;,
\label{eq1}
\end{equation}
where $\phi$ is the measured relative to the reaction plane. 
$v_2(p_T)$ rises strongly as a function of transverse momentum up to 
$p_T \approx 1.5\, \mbox{GeV}$.
One interpretation of 
the observed flow is that hydrodynamic pressure is built up from
the rescattering of produced secondaries  and
pressure gradients subsequently drive collective motion.
A strong  hydrodynamic response is possible 
if the sound
attenuation length $\Gamma_{s} \equiv  \frac{4}{3}\frac{\eta}{e + p}$,
is
significantly smaller than the expansion rate, $\sim \tau$. 
(In the formula $\Gamma_{s} \equiv  \frac{4}{3}\frac{\eta}{e + p}$,  $\eta$
is the shear viscosity, $e$ the energy density and $p$ the pressure.)
Estimates based upon
perturbation
theory give $\Gamma_{s} \sim \tau$ and indeed thirty times the perturbative 2-2
cross
sections are needed to obtain the  observed elliptic flow \cite{denes}.  
However, these
perturbative estimates are uncertain. 
In  an example of a  
strongly coupled gauge theory where calculations are possible (N=4 SUSY YM), 
$\Gamma_{s}$ is in fact approximately 2-4 times smaller 
compared to perturbation theory 
\cite{Son1} (see also Section \ref{Bjorken-intro}).

Ideal hydrodynamics ($\Gamma_{s}=0$)
has been used to simulate heavy ion reactions and readily reproduced the
observed
elliptic flow and its dependence on centrality, mass, beam energy and
transverse momentum \cite{Teaney1,Kolb1}. 
However ideal hydrodynamics failed in several respects. First, above
$p_T \approx 1.5 \, \mbox{GeV}$ 
the observed elliptic flow does not increase further
as predicted by hydrodynamics. Additionally, the 
single particle spectra deviate from  hydrodynamic predictions
above $p_{T} \approx 1.5\,\mbox{GeV}$.
Second, the observed HBT radii
are significantly smaller than predicted by ideal hydrodynamics 
\cite{BassDumitru,HBT-star,HBT-phenix}.  In 
particular, the longitudinal radius $R_{L}$ is 
50\% smaller than the ideal hydrodynamic result.
Further, the ratio between the outward 
($R_{O}$) and sideward ($R_{S}$) radii is observed to be 
approximately one while ideal hydrodynamics predicts 
$R_{O}/R_{S} \approx 1.3$ \cite{BassDumitru}. 

The domain of applicability of hydrodynamics can be answered
quantitatively by calculating the first viscous correction 
to ideal hydrodynamic results.
The effect of viscosity is twofold. First, viscosity changes
the solution to the equations of motion. Second, viscosity changes
the local thermal distribution function. This 
effect was first investigated in heavy ion physics by Dumitru \cite{dumitru}. 
The purpose of this work
is to consider  the effect of a modified thermal distribution function
on spectra, elliptic flow,  and HBT radii. Thus this work  delineates
the boundaries of the hydrodynamic description as applied to 
relativistic heavy ion collisions.

\section{Viscous Corrections to a Boost Invariant expansion}
\label{Bjorken-intro}

First consider a baryon free viscous boost invariant expansion 
with a vanishing bulk viscosity, 
but a non-zero shear viscosity, $\eta$. 
Note throughout this work we denote the space-time rapidity 
as $\eta_s$ and the viscosity as $\eta$. 
Unlike for ideal hydrodynamics where entropy is conserved, 
the entropy per unit space-time rapidity $\tau s$ increases as a function of $\tau=\sqrt{t^2 - z^2}$ \cite{Hwa,BJ,GBaym,MG84}
\begin{equation}
   \frac{d ( \tau s) }{d\tau} = 
   \frac{\frac{4}{3} \eta}{\tau T} \; .
\end{equation}
For hydrodynamics to be valid,  the entropy produced over the
time scale of the expansion $\tau$ (to wit, 
$\tau \frac{\frac{4}{3} \eta}{\tau T}$)  must be small compared to the 
the total entropy, ($\tau s$). This leads to the requirement that 
\begin{equation}
     \frac{\Gamma _{s}}{\tau} \ll 1 \; ,
\end{equation}
where we have defined the {\it sound attenuation length} 
\begin{equation} 
    \Gamma_{s} \equiv \frac{ \frac{4}{3} \eta } {sT} \;.
\end{equation}
$\Gamma_s$ is approximately the  mean free path 
and therefore  the condition $\Gamma_s/\tau \ll 1$ is 
just the statement that the mean free path be small 
compared to the system size.
The name ``sound attenuation length''  follows 
from the dispersion relation for a sound pulse  
$\omega = c_s k + \frac{1}{2}\, i \,\Gamma_s\,k^2$, where $c_s^2 =
 \left( \frac{\partial p}{\partial e}\right) $ is
the squared speed of sound.  
In the remainder of this section, I gather estimates for $\Gamma_s$ in
the Quark Gluon Plasma (QGP).
For similar estimates in the hadron gas see \cite{Raju-Reports}.

The shear viscosity has been determined in the perturbative QGP only to 
leading log accuracy \cite{Yaffe,GBaym-eta}. 
To leading $\log(g^{-1})$ the shear viscosity with two 
light flavor is given 
by $\eta  = 86.473 \,\frac{1}{g^4} \frac{T^{3}}{\log(g^{-1})}$. 
With  the entropy of  the QGP, $s=37\, \frac{\pi^2}{15} T^{3}$ and setting 
$\alpha_{s}\rightarrow\frac{1}{2}$ and $\log(g^{-1})\rightarrow 1$ 
the sound attenuation length in perturbation theory is  
\begin{eqnarray}
\left(\frac{\Gamma_{s}}{\tau}\right)_{Pert.}=0.18 \frac{1}{\tau T}  \; .
\end{eqnarray}
Estimates of evolution time scales give $\tau T \sim 1$.  
The value of $\Gamma_s/\tau$ is  sensitive to the value of 
$\alpha_s$. 

This perturbative estimate of $\Gamma_s$
is clearly uncertain and assumes that $\alpha_s\approx 1/2$ and that 
$\log(g^{-1})$ is 
a large number. Recently the shear viscosity was evaluated in 
a strongly coupled gauge theory, $N=4$ SUSY YM using the 
AdS/CFT correspondence \cite{Son1}.
The  shear viscosity is given 
by $\eta=\frac{\pi}{8} N^2_{c} T^3$ \cite{Son1} and the entropy is given by 
$s=\frac{\pi^2}{2} N^2_{c} T^3$ \cite{Gubser}. Thus in this  strongly
coupled field theory $\Gamma_s$ is 
\begin{equation}
     \left(\frac{\Gamma_s}{\tau}\right)_{AdS/CFT}  = \frac{1}{3 \pi \tau T} \; .
\end{equation}
which is 2-4 times smaller
than the corresponding perturbative estimate depending. 

Finally, I compare these theoretical estimates of $\Gamma_s$
to the value abstracted from Monte Carlo simulations of RHIC collisions  
performed by Gyulassy and Molnar (GM) \cite{denes}. GM modeled the 
heavy ion reaction as a gas of massless classical particles 
suffering only  $2\rightarrow2$  elastic collisions with a constant
cross section  in the c.m.s frame, 
$\frac{d\sigma}{d\Omega} = \frac{\sigma_{0}}{4\pi}$.  When particle
number is conserved,  $\Gamma_s$  is given 
by a more complicated formula  which 
reflects the 
coupling between the energy and number densities \cite{Weinberg}
\begin{widetext}
\begin{equation}
   \Gamma_s = \frac{\frac{4}{3}\eta}{e + p} + \frac{\kappa}{e+p} 
              \left(\frac{\partial e}{\partial T}\right)_n^{-1}
              \left[ e + p - 
              2 T \left(\frac{\partial p}{\partial T}\right)_n
               + c_s^2 T 
               \left(\frac{\partial e}{\partial T}\right)_n
               -  \frac{n}{c_s^2} 
               \left(\frac{\partial p}{\partial n}\right)_T 
               \right] \; ,
\end{equation}
\end{widetext}
where $\kappa$ is the
thermal conductivity.
For the GM gas,  
$c_s^2 = \frac{1}{3}$, $p = \frac{1}{3}e = nT$ and  
$\Gamma_s$ reduces to $\frac{ \frac{4}{3}\eta } {e + p}$ as before.
The shear viscosity in the GM gas is
$\eta \approx 1.264 \frac{T}{\sigma_0}$ \cite{deGroot-vis}.
Therefore $\Gamma_s$ is 
directly proportional to the mean free path 
\begin{eqnarray}
\Gamma_s = 0.421 \frac{1}{n \sigma_0}\;.
\end{eqnarray}
In order to achieve a reasonable agreement with the measured elliptic 
flow, GM required a 
transport opacity of $\chi \approx 20 \div 40$.  This 
transport opacity was reached when the cross section was
$\sigma_0 \approx 10\div20\,\mbox{mb}$  and the number 
of particles was $\frac{dN}{d\eta} \approx 1000$ at  
proper time $\tau_o = 0.1\,\mbox{fm}$. The initial density of 
particles is $n = \frac{dN}{d\eta}/(\tau_o \pi R^2)$. Substituting 
$R \approx 5.5\,\mbox{fm}$ we obtain 
\begin{eqnarray}
      \left(\frac{\Gamma_s}{\tau}\right)_{GM} &=&  0.02 \div 0.04  \; .
\end{eqnarray}
This is smaller by a factor of three or more than even the AdS/CFT estimate 
assuming that $\tau T \sim 1$. The physical mechanism for 
such a small viscosity remains unclear.


The sound attenuation length is uncertain. In what follows
we take $\frac{\Gamma_{s}}{\tau} = \frac{1}{5}$ and calculate
viscous corrections to the observed spectra, elliptic flow, and
HBT radii. In summary,  
perturbation theory finds $\Gamma_s/\tau \approx 0.18$,
strongly coupled supersymmetric field theory finds
$\Gamma_s/\tau \approx 0.11$, and 
phenomenology finds $\Gamma_s/\tau \approx 0.03$. 

\section{Viscous Corrections to the Distribution Function}

Viscosity  modifies the thermal distribution function. 
The formal procedure for determining the viscous corrections  
to the thermal distribution function is given in the references
\cite{Yaffe,deGroot}. In general, for a multi-component gas 
the viscous correction is different for each component. For
simplicity, we will consider a single component gas of ``pions'' 
with $m_{\pi} = 140\,\mbox{MeV}$.
The basic form of the viscous correction
can be intuited without calculation. First write 
$f(p) = f_{o}+ \delta f$, where $f_{o}(\frac{p\cdot u}{T})=\frac{1}{e^{p\cdot u/T} - 1}$ is the equilibrium
thermal distribution function and $\delta f$ is the 
first viscous correction.   
$\delta f$ is linearly proportional to the spatial gradients in the system. 
Spatial gradients which have no time derivatives in the rest frame  and
are therefore formed with the differential operator 
$\nabla_{\mu} = (g_{\mu\nu} - u_{\mu}u_{\nu})\partial^{\nu}$\,.
For a baryon free fluid, these gradients are $\nabla_{\alpha}T$,  
$\nabla_{\alpha}u^{\alpha}$, and 
$\left\langle \nabla_{\alpha}u_{\beta} \right\rangle$, where  
$\left\langle \nabla_{\alpha}u_{\beta} \right\rangle \equiv 
\nabla_{\alpha}u_{\beta} + \nabla_{\beta}u_{\alpha} - 
\frac{2}{3} \Delta_{\alpha\beta}\nabla_{\gamma}u^{\gamma}$.
$ 
\nabla_{\alpha}T$  can be converted into  spatial derivatives 
$\nabla_\alpha u_{\beta}$ using the ideal equations of 
motion and the condition that $T^{\mu\nu}u_{\nu} = e u^{\mu}$ \cite{deGroot}.
$\nabla_{\alpha}u^{\alpha}$ 
leads ultimately to a bulk viscosity and will be neglected in 
what follows. Finally, $\left\langle \nabla_{\alpha}u_{\beta} \right\rangle$ 
leads to a shear viscosity. If $\delta f/f_{o}$ is
restricted to be a polynomial of degree less than three in $p^{\mu}$, then the functional
form of the viscous correction  is completely determined,
\begin{eqnarray}
\label{correction}
   f = f_{o}\,(1 + \frac{C}{2 T^3} p^{\alpha}p^{\beta} 
   \left\langle \nabla_{\alpha}u_{\beta} \right\rangle)\;.
\end{eqnarray}   
For a Boltzmann gas this is the form of the viscous correction adopted
in this work. 
The factor of 2 in $\frac{C}{2 T^3}$ is inserted for later convenience.
For Bose and  Fermi
gasses the ideal distribution function in Eq. \ref{correction}  
is replaced with $f_{o}(1 \pm f_{o})$ \cite{Yaffe}.
The correction described here is precisely the ``first approximation''
of reference \cite{deGroot} and the  ``one parameter ansatz''  for
a variational solution of reference \cite{Yaffe}. The ``one parameter ansatz'' 
reproduces the full result to the $15\%$ level.

The coefficient $C$ in Eq. \ref{correction} can be reexpressed in terms
of the sound attenuation length.  Indeed, substituting 
$f$ to determine the stress energy tensor 
\begin{equation}
  T^{\mu\nu} =  T^{\mu\nu}_{o} + \eta \left\langle \nabla^{\mu}u^{\nu} \right\rangle  
   =  \int \frac{d^3p}{(2 \pi)^3 E} p^{\mu}p^{\nu} f \; ,
\end{equation}
we find
\begin{widetext}
\begin{equation}
\label{visrelation}
   \eta \left\langle \nabla^{\mu}u^{\nu} \right\rangle  
   = \frac{C}{2 T^3}\left[ \int \frac{d^3p}{(2 \pi)^3 E} p^{\mu}p^{\nu} p^{\alpha} p^{\beta}
       f_{o}(1 + f_{o})\right] 
       \left\langle \nabla_{\alpha}u_{\beta} \right\rangle  \; .
\end{equation}
\end{widetext}
The quantity in square brackets is a fourth rank symmetric tensor and
consequently can be written in terms of 
$\Delta^{\mu\nu}\equiv g^{\mu\nu} - u^{\mu}u^{\nu}$ and  
$u^{\mu}$. Thus,
\begin{widetext}
\begin{eqnarray}
\label{tensor}
\frac{C}{2 T^3}
\int \frac{d^3p}{(2 \pi)^3 E} p^{\mu}p^{\nu} p^{\alpha} p^{\beta} 
f_o (1 + f_o)
= a_{o}  \left(u^{\mu} u^{\nu} u^{\alpha} u^{\beta}\right)  + a_{1} 
\left( \Delta^{\mu\nu} u^{\alpha} u^{\beta} + \mbox{permutations} \right) \\ \nonumber
+ a_{2} \left( \Delta^{\mu\nu} \Delta^{\alpha \beta} + 
\Delta^{\mu\alpha}\Delta^{\nu\beta} + \Delta^{\mu\beta}\Delta^{\nu\alpha} \right) \;.
\end{eqnarray}
\end{widetext}
Substituting Eq.\,\ref{tensor} into Eq.\,\ref{visrelation} and 
using the identities 
$u^{\alpha}\left\langle \nabla_{\alpha}u_{\beta} \right\rangle=
u^{\beta}\left\langle \nabla_{\alpha}u_{\beta} \right\rangle=
\Delta^{\alpha\beta}\left\langle \nabla_{\alpha}u_{\beta} \right\rangle=0$,
we find $2 a_{2} = \eta$. To determine the coefficient $a_{2}$,  contract
both sides of Eq.\,\ref{tensor} with 
\begin{equation}
\frac{1}{45} \left( \Delta^{\mu\nu} \Delta^{\alpha\beta} + 
\Delta^{\mu\alpha}\Delta^{\nu\beta} + \Delta^{\mu\beta}\Delta^{\nu\alpha} \right) \; ,
\end{equation}
and evaluate the resulting expression in the local rest frame. The
result for the viscosity is
\begin{equation}
\label{etaequation}
  \eta = \frac{6}{90} \frac{C}{T^3} \int \frac{d^3p}{(2\pi)^3 E} 
         f_{o} ( 1 + f_{o} )\, {|{\bf p}|}^{4} \;.
\end{equation}
For a Boltzmann gas $f_o (1+f_o)$ is 
be replaced with 
$f_o(\frac{p\cdot u}{T}) = e^{-\frac{p\cdot u}{T}}$ and the
integrals can be performed analytically.
Comparing the resulting expression to the entropy of an ideal Boltzmann gas (see e.g. \cite{Raju})
we find  $C=\frac{\eta}{s}$.  
For a massless Bose 
gas the integrals can again be performed analytically and 
$C=\frac{\pi^4}{90\zeta(5)} \frac{\eta}{s} \approx 1.04 \frac{\eta}{s}$. 
For a massive
Bose gas, the integral was performed numerically and  
$C$ varies monotonously between these  two limiting cases.
Therefore up to a few percent, we have 
$C = \frac{\eta}{s}$, and the viscous correction $\delta f$ is
\begin{eqnarray*}
\label{deltaf}
   \delta f &=& \frac{3}{8} \frac{\Gamma_s}{T^2} \,f_{o}(1 + f_{0}) \,
   p^{\alpha}p^{\beta} 
   \left\langle \nabla_{\alpha}u_{\beta} \right\rangle\;.
\end{eqnarray*}   
      
\section{Viscous Corrections to a Bjorken Expansion}
\label{BjorkenSect}

Before considering the viscous corrections to more general 
hydrodynamic expansions,
let us consider a simple Bjorken expansion of
infinitely large nuclei without transverse flow.   
At mid space-time rapidity the stress energy tensor is 
at time $\tau_o$ is given by \cite{MG84}
\begin{eqnarray}
 T^{\mu\nu}_{o} +  \eta \left\langle \nabla^{\mu}u^{\nu} \right\rangle &=&  
 \bordermatrix{ & t & x & y & z \cr
  t& e & 0 & 0 & 0 \cr
  x& 0 & p +\frac{2}{3}\frac{\eta}{\tau_o} & 0 & 0 \cr
  y& 0 & 0 & p + \frac{2}{3}\frac{\eta}{\tau_o} & 0  \cr
  z& 0 & 0 & 0  & p - \frac{4}{3}\frac{\eta}{\tau_o} \cr} \, ,
\end{eqnarray}
where, $T^{\mu\nu}_{o}$ denotes the ideal stress energy tensor
$\mbox{diag}(e, p, p,p)$,  
Thus, 
the longitudinal pressure  
is reduced  by the expansion,
$T^{zz} = p - \frac{4}{3}\frac{\eta}{\tau_{o}}$,
while the transverse pressure 
is  increased by the expansion,   
$T^{xx} = p + \frac{2}{3}\frac{\eta}{\tau_{o}}$.

The difference between the longitudinal and transverse pressures is
reflected in the $p_{T}$ spectrum of thermal distribution.  Since
the transverse pressure ($T^{xx}$) is increased by 
$\frac{2}{3}\frac{\eta}{\tau_{o}}$,
the particles are pushed out to larger $p_{T}$. 
Armed with the modified thermal distribution function, 
the Cooper Frye formula \cite{Cooper-Frye}  
gives the  thermal spectrum  of particles
in the transverse plane at proper time  $\tau_o$
\begin{subequations}
\label{CooperFrye}
\begin{eqnarray}
    \frac{d^{2}N}{d^{2}p_{T}\,dy} &=&  
    \frac{1}{(2\pi)^3} \int p^{\mu} d\Sigma_{\mu} \, f   \\
    \frac{d^{2}N^{(0)}}{d^{2}p_{T}\,dy} + 
    \frac{d^{2}N^{(1)}}{d^2p_T\,dy}  &=&  \frac{1}{(2\pi)^3} 
    \int p^{\mu} d\Sigma_{\mu}\, f_{o}  + \delta f \;. 
\end{eqnarray} 
\end{subequations}
Here $d\Sigma_{\mu}$ is the oriented space-time volume.
Substituting into   
Eq. \ref{CooperFrye} (see Appendix \ref{BAppend})
we obtain the 
the ratio between the viscous correction 
($\delta\, dN\equiv \frac{dN^{(1)}}{d^2p_T dy}$) and the
ideal spectrum ($dN^{(0)} \equiv \frac{dN^{(0)}}{d^2p_T dy}$)
\begin{eqnarray*}
\label{spectra}
  \frac{\delta\,dN}{ dN^{(0)} }  &=& \frac{\Gamma_s} {4\tau_o}  
  \left\{    \left( \frac{p_T}{T}
             \right)^2  - 
             \left( \frac{m_T}{T} 
             \right)^2 
             \frac{1}{2} 
             \left( \frac{
                           K_3(\frac{m_T}{T})
                         }{ 
                           K_1(\frac{m_T}{T}) 
                         } -1
             \right) 
  \right\}\;.
\end{eqnarray*}
Using the asymptotic expansion for the modified Bessel functions,  we have 
for large transverse momenta, 
\begin{eqnarray}
\label{spectra2}
\frac{\delta\,dN}{ dN^{(0)} } =  
\frac{\Gamma_s}{4\tau_o} \left( \frac{p_{T}}{T} \right)^2.  
\end{eqnarray}
As promised, the larger transverse pressure drives pushes the 
corrected spectrum out to higher transverse momenta.
For a Bjorken expansion without transverse flow, this formula also 
indicates at what transverse momentum the 
hydrodynamic description of $p_T$ spectra  is applicable. 
For $\frac{\Gamma_{s}}{\tau_o} \approx \frac{1}{5}$, and $T=200\;\mbox{MeV}$ the
ratio between the ideal spectrum and the correction becomes of order
one for $p_{T}^{max} \approx 800\;\mbox{MeV}$. We shall see in 
the next section that this  upper bound on the domain
of hydrodynamics is significantly larger $p_{T}^{max}\approx 1.5\;\mbox{GeV}$
once the transverse expansion is included in the flow profile.

\begin{figure}
\includegraphics[height=3.2in,width=3.2in]{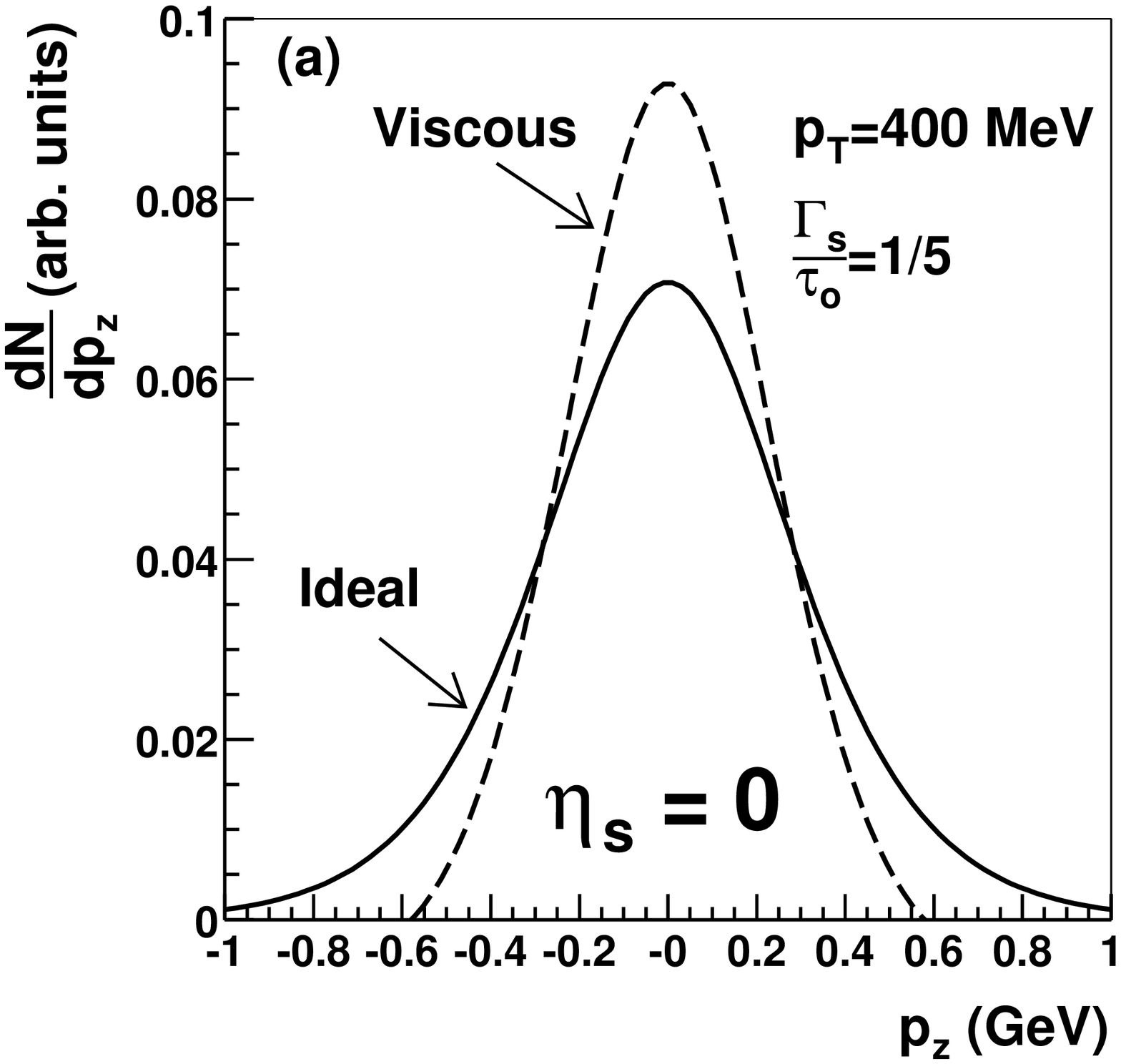}
\includegraphics[height=3.2in,width=3.2in]{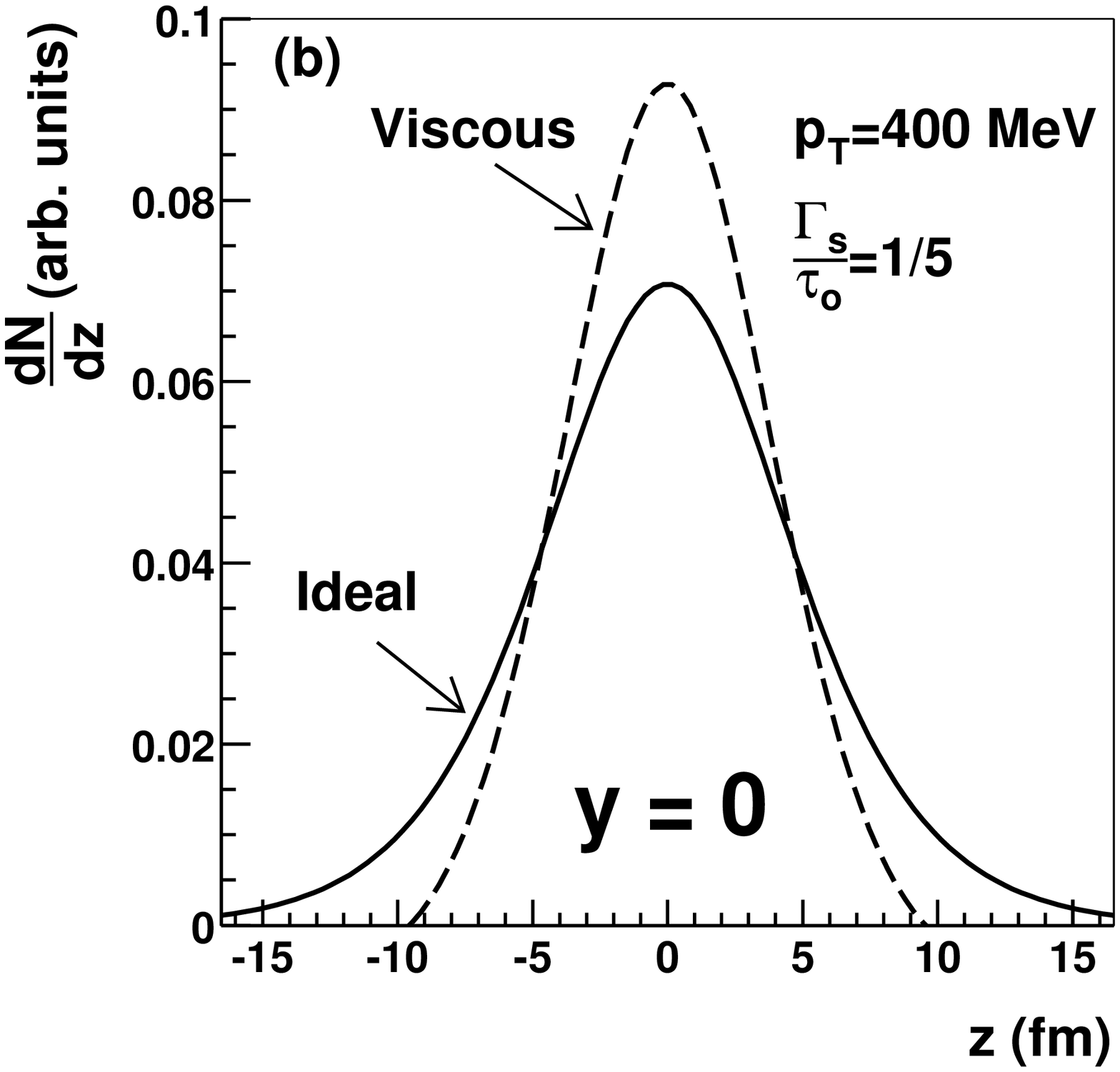}
\caption{(a) The $p_z$ distribution of particles 
with coordinate-space
rapidity $\eta_s=0$, with and without viscous corrections.
(b) The $z$ distribution of particles  
with momentum-space rapidity 
$y=0$, with and without viscous corrections.  
The curves are drawn for a Bjorken expansion without
transverse flow at $\tau_o=7\,\mbox{fm}$ for a Boltzmann gas 
with temperature, $T=160\,\mbox{MeV}$, $m=140\,\mbox{MeV}$. The transverse
momentum is fixed, $p_T=400\,\mbox{MeV}$.  The viscous 
correction is linearly proportional to $\Gamma_s/\tau_o$.
}
\label{fighbta}
\end{figure}

We have already noted that the longitudinal pressure is 
reduced by the expansion, $T^{zz} = p - \frac{4}{3}\frac{\eta}{\tau}$. 
The reduction in the longitudinal pressure is ultimately 
responsible for a reduction in the longitudinal radius measured
by Hanbury-Brown Twiss interferometry. 
Since the longitudinal pressure 
is reduced due to the expansion, the distribution in $p_{z}$ at 
mid space-time rapidity ($\eta_{s}=0$) is 
narrower.   This is illustrated  
in Fig. \ref{fighbta}(a) for a fixed transverse momentum $p_{T}=400\,\mbox{MeV}$. 
Due to boost invariance  the $p_{z}$ distribution at $\eta_{s}=0$ 
is directly related to the $z$ distribution at $y=0$ \cite{GBaym}.
Specifically, for fixed transverse momentum, 
$\frac{dN}{dy d\eta_s}$ is a function of $\left|y -\eta_{s}\right|$, 
which leads to the relation 
\begin{equation}
   \left. m_{T} \frac{dN}{dp_{z} d\eta}\right|_{\eta=0} 
   = \left.\tau_{o} \frac{dN}{dy dz} \right|_{y=0}.
\end{equation}
It follows that the z distribution at mid momentum-space rapidity 
is narrower as indicated in 
Fig. \ref{fighbta}(b). The width of 
this z-distribution, 
is related to the longitudinal radius that
is measured by HBT interferometry (see e.g. \cite{Wiedemann}). 

To understand this result analytically we must calculate 
the width of z distribution for a simple Bjorken expansion of 
a Boltzmann gas at proper time $\tau_o$.  
Let us
quickly recall the definitions  of the HBT radii.  
The source function $S(x,K)$ for on shell pion emission is defined such that
\begin{equation}
\label{source}
E_{K} \frac{d^3N}{d^3K } \equiv \int d^4x \, S(x, K)
\end{equation}
where $E_K= K^{0} = \sqrt{{\bf K}^2 +  m_{\pi}^2}$. 
Averages with respect to the source function are 
 defined as $ \left\langle\alpha\right\rangle_{\bf K} \equiv \int d^4x\, \alpha \,S(x,K)/ \int d^4x\, S(x,K) $. 
To a good approximation 
(see e.g. Ref \cite{Wiedemann}), 
certain spatial and temporal variances of the 
source function can be determined from 
the Bose-Einstein correlations between pion pairs at small relative momenta.
For a boost invariant and rotationally invariant source, we can 
assume without loss of generality that 
the pair momentum  points in the x direction (i.e.  
${\bf K}=(K^{x}, K^{y}, K^{z})=(K_{T},0,0)$). Then 
the following variances can be determined from HBT measurements
\begin{eqnarray}
     R_{O}^2 (K_{T}) &\equiv& \left\langle (\tilde{x}- v_{K}\tilde{t})^2 \right\rangle_{K_T}   \\
     R_{S}^2 (K_{T}) &\equiv& \left\langle \tilde{y}^2 \right\rangle_{K_T}  \\
     R_{L}^2 (K_{T}) &\equiv& \left\langle \tilde{z}^2 \right\rangle_{K_T} \; , 
\end{eqnarray}
where $v_{K} = K_{T}/E_K$  and  for example 
$\tilde{x} \equiv x - \langle x\rangle$.
Comparing Eq. \ref{spectra} and Eq. \ref{source},  we see that in this 
work the source function  is confined to a freezeout surface
and therefore the averages are understood to mean
\begin{equation}
\label{fave}
   \left\langle\alpha\right\rangle_{\bf K} \equiv 
   \frac{
      \int_{\Sigma} K^{\mu}\,d\Sigma_{\mu}\, \alpha \,f(x,K) 
   }{
      \int_{\Sigma} K^{\mu}\, d\Sigma_{\mu}\, f(x,K)  
   } \; .
\end{equation}

The assumption of a sharp freezeout surface is clearly unrealistic. 
In general there is a  transition region from  hydrodynamics
to the Knudsen limit.
Within ideal hydrodynamics this transition region can not be determined. 
Within viscous hydrodynamics, viscous terms 
become large ($\sim 1/2$) and signal the  
transition.

Armed with these formula,  the computation of $R_{L}^2$ for a 
boost invariant expansion is straight forward. We have
\begin{equation}
\label{CFBj}
     R_{L}^2 (K_{T}) \equiv \left\langle \tilde{z}^2 \right\rangle_{K_T}  
    \equiv \frac{
    \int K^{\mu} d\Sigma_{\mu}\, f(x,K)\,  z^2 
    }{
    \int K^{\mu} d\Sigma_{\mu}\, f(x,K) 
    } \;.
\end{equation}
Substituting $f = f_{o} + \delta f$, expanding to 
first order  in $\delta f$, and performing the integrals (see Appendix \ref{BAppend})
we find  the viscous correction $\delta R_{L}^2$ 
\begin{equation}
\label{HBTBj}
   \frac{
          \delta R_L^2
        }{
          (R_L^2)^{(0)}
        }  =  
-\frac{\Gamma_s}{\tau_o} 
\left[
       \frac{6}{4} \frac{m_T}{T} 
      \frac{ 
         K_3(\frac{m_T}{T})
      }{ 
         K_2(\frac{m_T}{T})
      }  -
      \left(\frac{m_T}{T} \right)^2 \frac{1}{8} 
      \left( \frac{
                    K_3(\frac{m_T}{T})
                  } {
                    K_2(\frac{m_T}{T})
                  } - 1
      \right) 
\right] \; ,
\end{equation}
where the $(R_{L}^2)^{(0)}$ is the ideal longitudinal radius \cite{Bertsch}
\begin{equation}
\label{HBTideal}
   (R_L^2)^{(0)} = \tau_{o}^2 \frac{T}{m_T} \frac{K_2(x)}{K_1(x)} \; .
\end{equation}
For the relevant range of $\frac{m_{T}}{T}$, the Bessel function expression in square brackets is large $\approx 6-8$. Accordingly, viscous corrections
to the longitudinal radius are quite large ($> 100\%$) 
and tend to reduce the radius relative
to its ideal value. Including the transverse expansion 
reduces the viscous correction to $50\%$ .
 Nevertheless, the viscous correction to 
the longitudinal radius remains large unless $\Gamma_s/\tau_o$ is
significantly smaller than $0.1$\;. This 
formula and some caveats are discussed further in  the next 
section.

\section{Viscous Corrections with Transverse Expansion}

To go further and illustrate the effect of viscosity on the observed spectra,
elliptic flow and HBT radii of hydrodynamical models of the 
heavy ion collision, 
I generalize the blast wave model to include 
the viscous corrections of  Eq.\,\ref{correction}.     
The blast wave
model provides of a simple parametrization of the flow
of full ideal hydrodynamic simulations which assume boost
invariance \cite{Kolb1,Teaney1}. The corrections 
described below are therefore indicative of similar corrections to 
these simulations. This is the reason
for adopting the blast wave model here.
The blast wave model also has been used to fit experimental 
data. The model provides a good description of 
spectra and elliptic flow \cite{Kolb1,v2-star2,Jane} and provides a fair 
description of HBT radii for small $M_T$, 
$M_{T} < 0.5\,\mbox{GeV}$ \cite{StarQM2002}. However, for larger
$M_T$ the model does not reproduce the strong dependence on 
$M_T$ seen in the $R_O$ and $R_S$ radii \cite{PhenixQM2002,Jane}. 
The blast wave model
remains simply a model of the flow fields and ultimately a full viscous 
simulation is needed to estimate viscous effects. 

In the blast wave model of central collisions considered here, a hot pion gas 
is expanding in a boost invariant fashion and freezes out at 
a proper time $\tau_{o}$. In the transverse 
plane, the temperature is constant $T_{o}=160\,\mbox{MeV}$
and the matter distribution is uniform up to a radius $R_{o}$. 
The transverse velocity  rises linearly as a function of the 
radius, $u^{r} = u^{o} \frac{r}{R_{O}}$.  
Summarizing, the hydrodynamic
fields ($T$ and $u^{\mu}$) are parameterized as
\begin{subequations}
\label{flowa}
\begin{eqnarray}
     T(\tau_o, \eta_{s}, r, \phi) &=&T_{o} \, \Theta (R_{o} - r)  \\
     u^{r}(\tau_o,\eta_s, r, \phi) &=& 
                        u_{o} \frac{r}{R_{o}} \, \Theta (R_{o}-r) \\
     u^{\phi} &=& 0 \\
     u^{\eta} &=& 0 \\
     u^{\tau_o} &=& \sqrt{1 + (u^{r})^2} \; .
\end{eqnarray}
\end{subequations}
The blast wave parameters are adjusted so that
model with the ideal thermal distribution 
can approximately reproduce the spectra and HBT radii.
Similar blast wave model fits have appeared ubiquitously in the 
heavy ion literature (see e.g. \cite{Jane}).
Then with the model parameters fixed, the viscous correction 
is calculated and compared to the ideal results. 
The 
model parameters for central collisions are 
recorded in Table\,\ref{blast-params}.
\begin{table} 
\begin{tabular} {|c|c|c|}  \hline
                   & \,Central (0-5\%)\, &\,Non-central (16-24\%)\,  \\ \hline\hline
\,$T_o$ (MeV)\,    &    160           &  160         \\ \hline
 $R_o$ (fm)        &    10            &    7.5          \\ \hline
 $\tau_{o}$ (fm)   &    7.0           &    5.25          \\ \hline
 $u_o$             &    0.55          &    0.55        \\ \hline
 $u_2$             &    0             &  0.1         \\ \hline
\end{tabular}
\caption{Table of parameters used in the blast wave model described in 
the text.}
\label{blast-params}
\end{table}

With the hydrodynamic fields specified,
the viscous tensor 
$\left\langle \nabla^{\alpha}u^{\beta} \right\rangle$
can be computed in a simple but lengthy calculation which 
is worked out in  
Appendix \ref{differentiate}. 
One technical point should be noted. In the viscous tensor 
$\left\langle \nabla^{\alpha}u^{\beta} \right\rangle $ time 
derivatives of the velocity appear. These time
derivatives are converted into spatial derivatives using 
the ideal equations of motion which are sufficient to leading order
in the viscosity.   

\begin{figure}[t!]
\includegraphics[height=3.2in,width=3.2in]{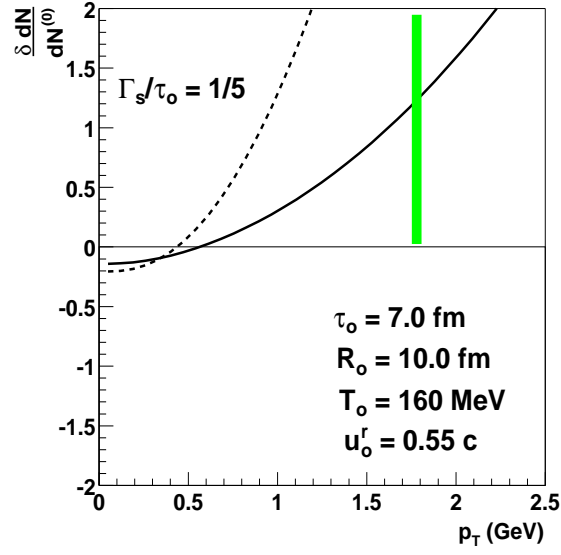}
\caption{
The solid line shows the ratio between the viscous correction 
($\delta\, dN\equiv \frac{dN^{(1)}}{d^2p_T dy}$) and the
ideal spectrum ($dN^{(0)} \equiv \frac{dN^{(0)}}{d^2p_T dy}$).
The 
dashed line shows the Bjorken result  without transverse 
flow given in Eq.\,\ref{spectra2}. 
The band indicates where the hydrodynamic description  of the
$p_T$ spectrum in the blast wave model can not be reliably 
calculated. The viscous correction
is linearly proportional to $\Gamma_s/\tau_o$.} 
\label{figSpectra}
\end{figure}

The spectrum of particles emerging from the freezeout oriented 3-volume
is calculated by employing the Cooper-Frye formula, Eq. \ref{CooperFrye}.
These integrals are performed numerically in a straightforward 
fashion. Again relevant details are relegated to  Appendix \ref{differentiate}. 
The ideal spectrum  of this blast wave model is typical of blast 
and is in rough agreement with pion data at RHIC. (See e.g. \cite{Jane} for 
fits to data of this type.) 
In Fig. \ref{figSpectra},  the solid line  shows the ratio of 
the viscous correction to the ideal spectrum. 
The dashed line shows the Bjorken result (Eq. \ref{spectra2}) without transverse flow. 
The viscous correction  becomes comparable to ideal
results for $p_T \approx 1.7\,\mbox{GeV}$ indicating the
breakdown of the hydrodynamic description of $p_T$ spectra for
the flow profile considered here. Setting $\Gamma_s/\tau_o$ to $0.1$ 
extends the domain of applicability to $2.3\,\mbox{GeV}$. The analytic Bjorken
result (Eq. \ref{spectra})
qualitatively explains the shape of  Fig. \ref{figSpectra}. 
Quantitatively however, the 
transverse expansion alleviates some of the longitudinal shear and
pushes the region of applicability hydrodynamics to somewhat 
larger transverse momentum.

Indeed, viscous effects 
are implicated in the heavy ion data for $p_{T} \approx 1.5\,\mbox{GeV}$. 
The observed
elliptic flow deviates from ideal hydrodynamic results for 
$p_{T} \approx
1.5\,\mbox{GeV}$.  
Further for $p_T \approx 1.5\,\mbox{GeV}$, 
the single particle spectra start to deviate 
strongly from the hydrodynamic results (see e.g. \cite{Teaney1}).  
Viscosity provides a simple explanation for the 
observed breakdown of the $p_T$ spectrum in this momentum range.

Next we examine the effect of viscosity on  elliptic flow. 
In non-central collisions
the radial velocity is
given a small elliptic component to reproduce the 
observed elliptic flow 
\begin{eqnarray}
   u^{r}(\tau_o,\eta_s, r, \phi) &=& 
                        u_{o} \frac{r}{R_{o}} (1 + u_2 \cos(2\phi)) \, \Theta (R_{o}-r) \; .
\end{eqnarray}
The functional form of  all other hydrodynamic 
fields is kept the same. Here we simulate the STAR 16-24\% centrality 
bin which corresponds to an impact parameter bin $\langle b\rangle\approx 6.8\;\mbox{fm}$ \cite{v2-star1}.
In the model, the radius and lifetime 
parameters ($R_o$ and $\tau_o$) are scaled downward from 
the central values
by the ratio of the r.m.s. radii
between $b=6.8\,\mbox{fm}$ and central AuAu collisions. 
This scaling of $R_o$ and
$\tau_o$
approximates the impact parameter dependence of ideal hydrodynamic 
solutions \cite{Teaney1}. The non-central parameters are recorded
in Table\,\ref{blast-params}.
As before, once the flow fields are specified, the viscous correction is found  
by differentiating 
$\left\langle \nabla^{\alpha} u^{\beta} \right\rangle$. The
full form of the correction is given in Appendix \ref{differentiate}.

The elliptic flow as a function
of transverse momentum $v_{2}(p_{T})$ is defined by Eq.~\ref{eq1}. 
 Expanding to first order 
\begin{eqnarray}
   v_{2}(p_{T}) &=& v_{2}^{ (0) }(p_{T})
   \left(1 - 
   \frac{
     \int d\phi  \frac{ d^2N^{(1)} } { p_T \,dp_{T} \,d\phi } 
   }{
     \int d\phi \frac{ d^2N^{(0)} } { p_{T}\,dp_{T}\, d\phi } 
   } 
   \right)
   + 
   \frac{ 
     \int d\phi \cos(2 \phi) \frac{ d^2N^{(1)} } { p_T \,dp_{T} \,d\phi } 
   }{ 
     \int d\phi \frac{ d^2N^{(0)} } { p_{T}\,dp_{T}\, d\phi } 
   } \; ,
\end{eqnarray}
where $v_2^{(0)}(p_T)$ denotes the elliptic flow 
as a function of $p_{T}$ calculated as in Eq.~\ref{eq1} but 
with the ideal distribution $\frac{dN^{(0)}}{p_T\, dp_T\, d\phi}$.

\begin{figure}
\includegraphics[height=3.2in,width=3.2in]{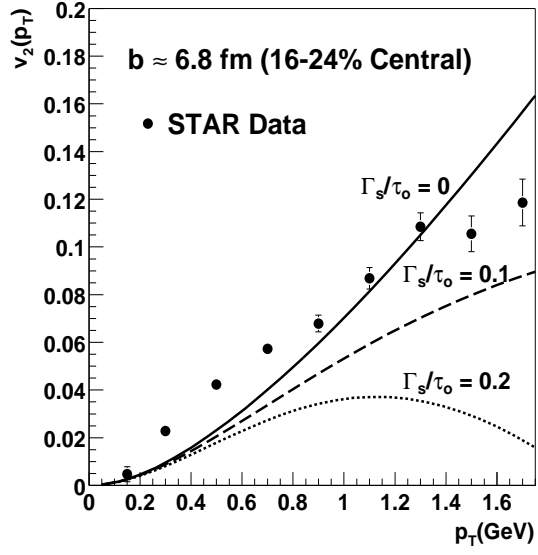}
\caption{Elliptic flow $v_2$ as a function of $p_T$ for different 
values of $\Gamma_s/\tau_{o}$. The data points 
are four particle cummulant data from the STAR collaboration \cite{v2-star1}.
Only statistical errors are shown. The difference between the 
ideal and viscous curves is linearly proportional to $\Gamma_s/\tau_o$.
}
\label{figv2}
\end{figure}

Fig. \ref{figv2}  shows the elliptic flow  for pions. By construction, 
the ideal curve $v_{2}^{(0)}$ roughly reproduces  the experimental
elliptic flow  at $b\approx 6.8\,\mbox{fm}$. Taking a more 
realistic flow profile would improve the agreement of the 
ideal results with data \cite{Kolb1}. 
The effect of viscosity 
is to reduce the elliptic flow. Similar 
results were recently found \cite{Heinz-Wong} by considering a 
partially thermalised expansion.  Taken at face value these 
results suggest that the viscosity is small.  Indeed,  
in order to agree with the ideal results up to $p_{T} \approx 1.0 \,GeV$ 
we require $\Gamma_s/\tau_o \lesssim 0.1$\,.  It must be mentioned that
the results of Fig. \ref{figv2}
are sensitive to the blast wave parameters. Ideal hydrodynamics 
generates an appropriate set of parameters. Whether a viscous expansion
(with $\Gamma_s/\tau_o= 0.1$) can reproduce the observed a 
elliptic flow remains an open question.

\begin{figure}
\includegraphics[height=3.2in,width=3.2in]{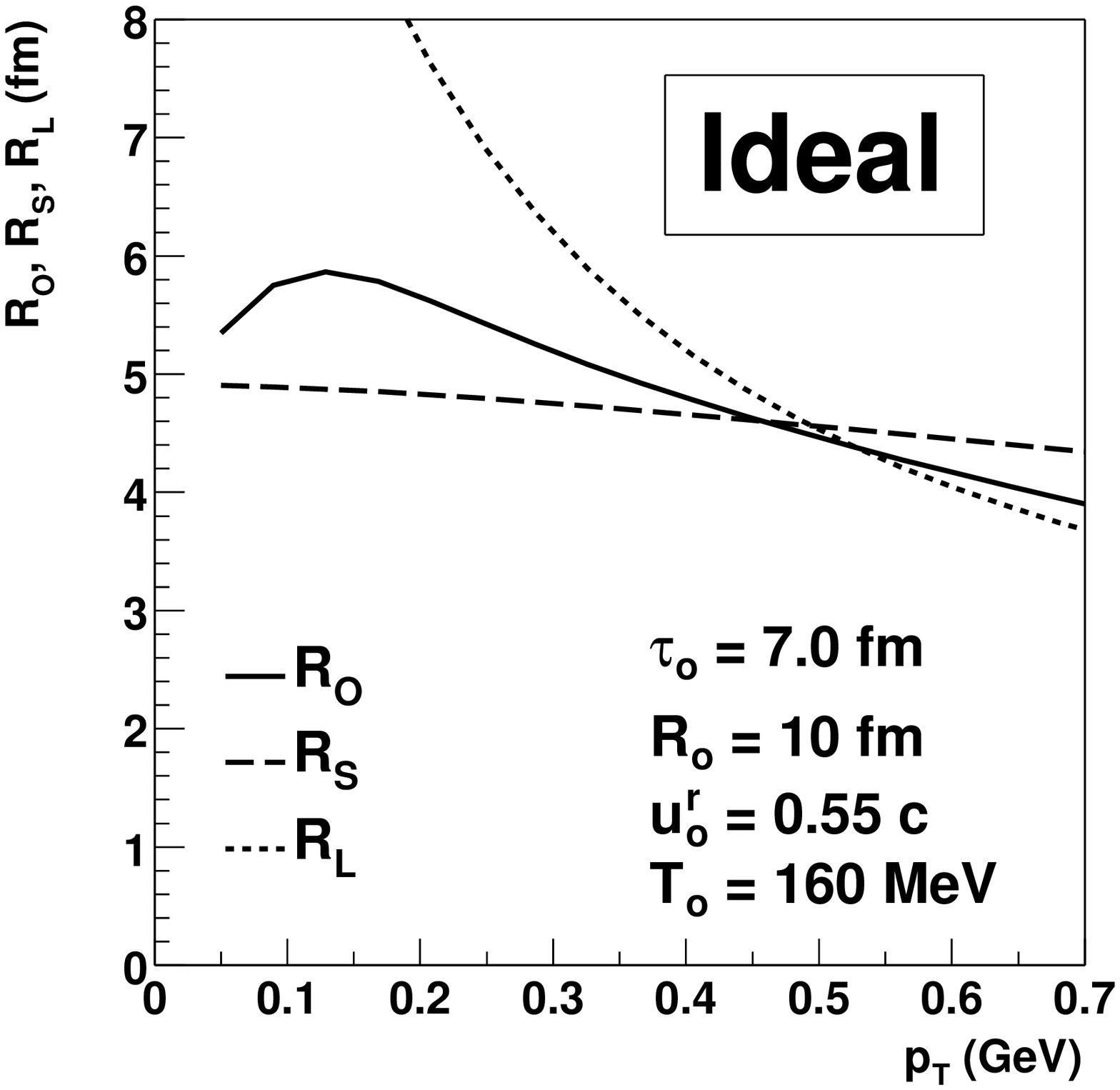}
\includegraphics[height=3.2in,width=3.2in]{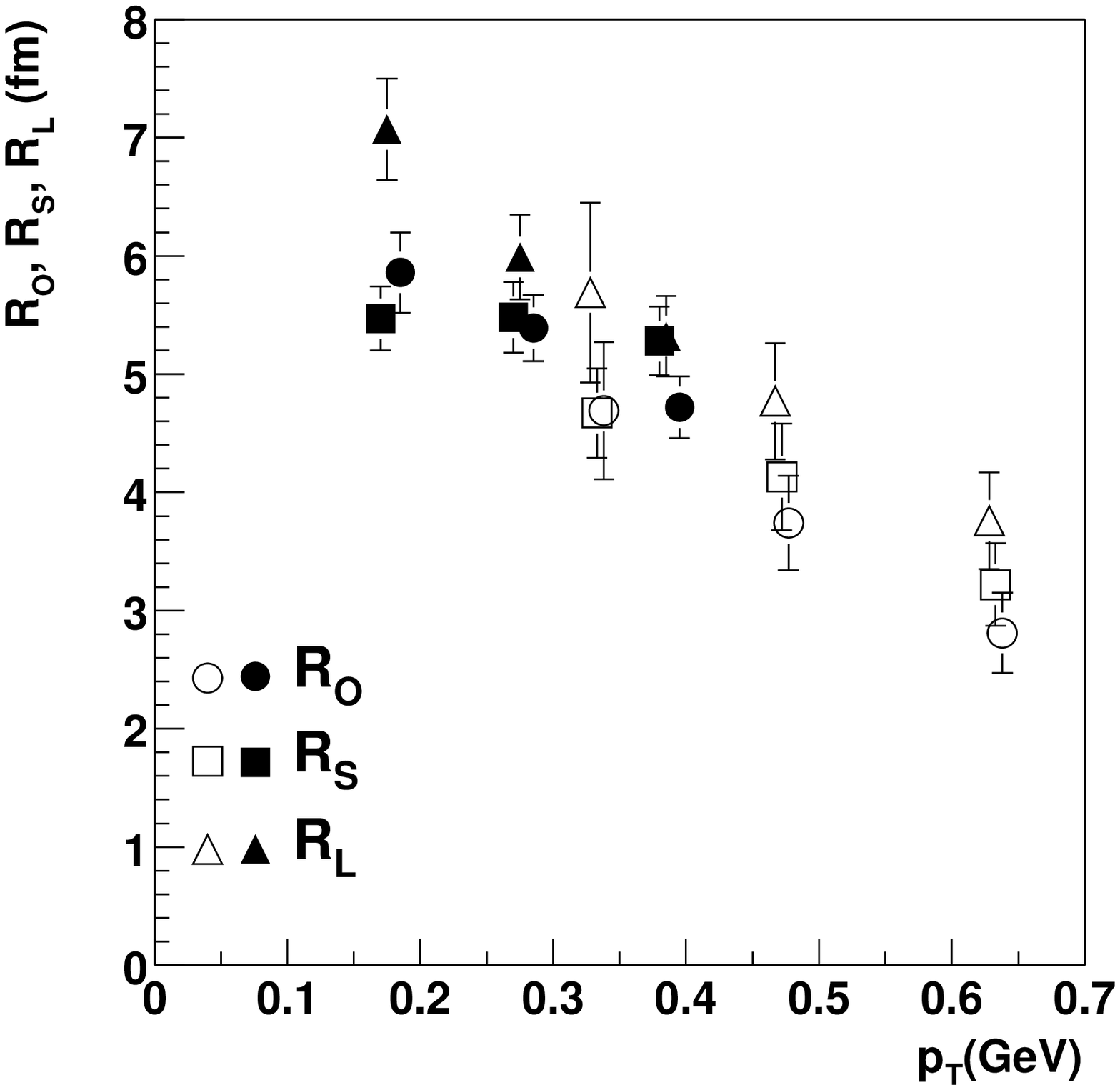}
\caption{(a) 
Ideal blast wave fit to  the 
experimental HBT radii $R_O$, $R_S$, and $R_L$  shown in (b) as 
a function of transverse momentum $K_T$. 
The solid symbols are 
from the STAR collaboration \cite{HBT-star} and the open symbols are from 
the PHENIX collaboration \cite{HBT-phenix}. For clarity, the experimental points
have been slightly shifted horizontally.
}
\label{idealhbt}
\end{figure}

\begin{figure}
\includegraphics[height=3.2in,width=3.2in]{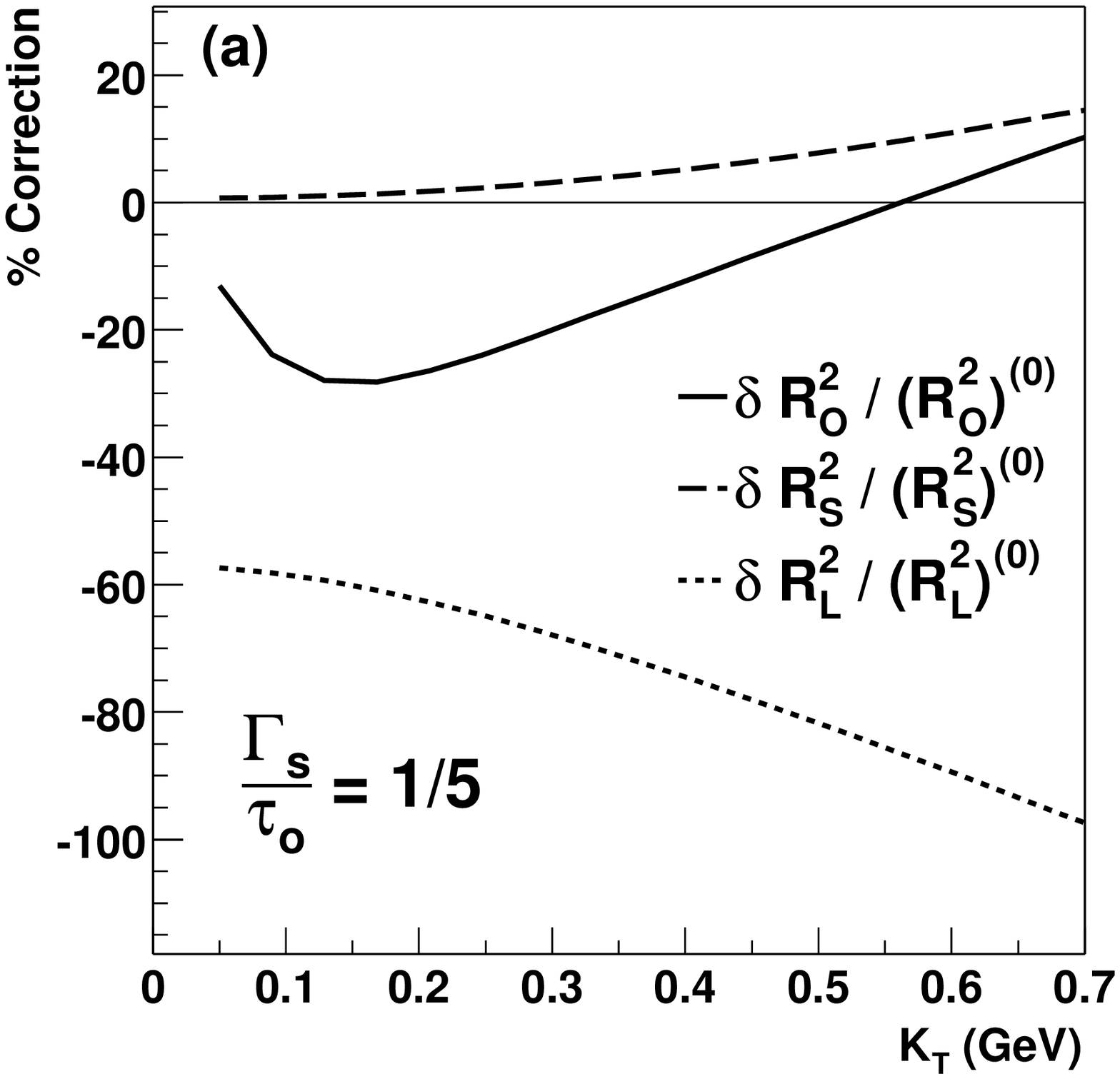}
\includegraphics[height=3.2in,width=3.2in]{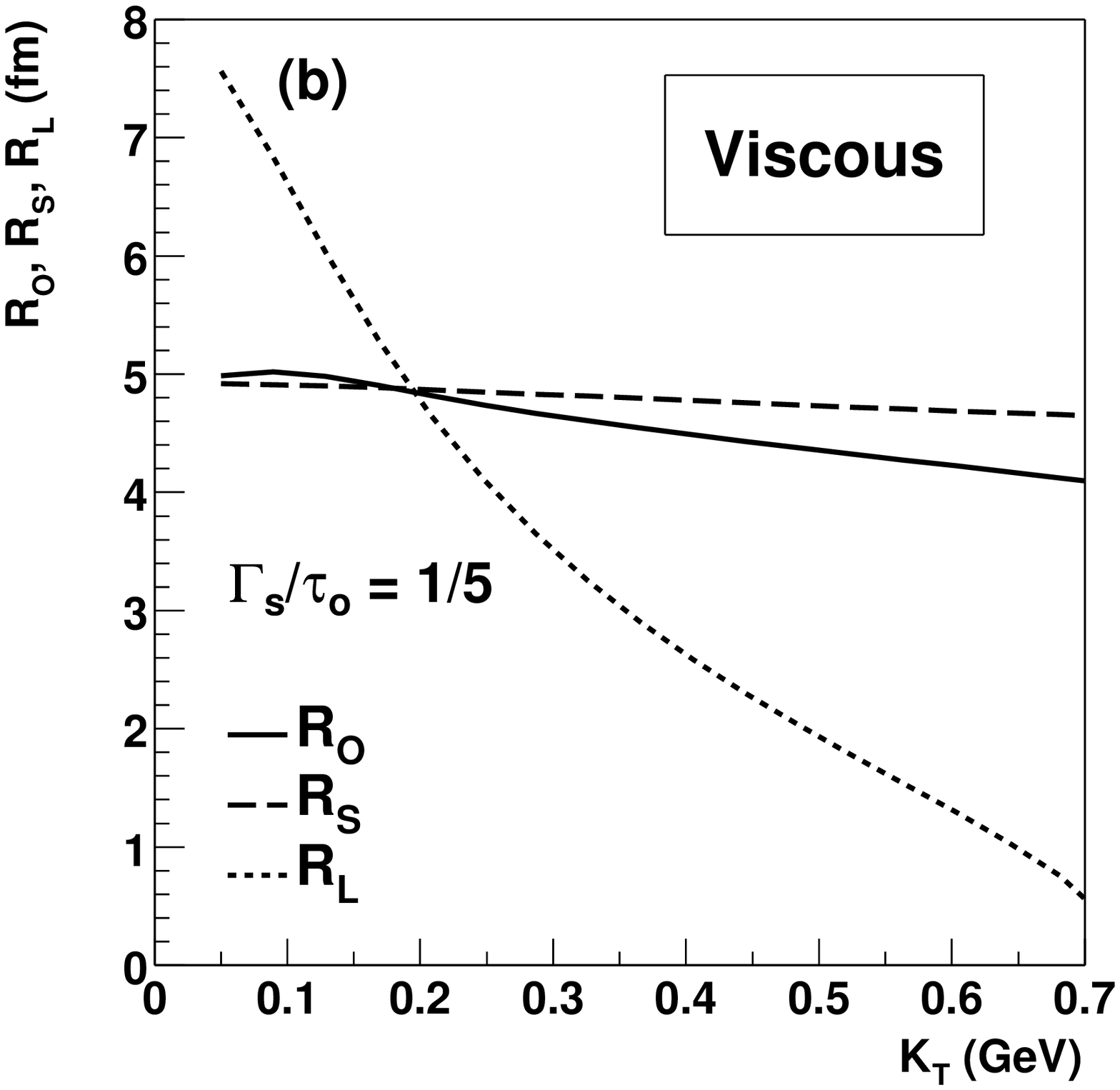}
\caption{
(a) Viscous correction $\delta R^2$  
for $R_O$, $R_S$, and $R_L$
relative to ideal blast wave HBT radii $(R^2)^{(0)}$ .
(b) 
The HBT radii $R_{O}$, $R_{S}$, and $R_L$ including
the viscous correction.
The viscous 
correction is linearly proportional to $\Gamma_s/\tau_{o}$.
}
\label{vishbt}
\end{figure}

Finally, I discuss how viscosity effects HBT radii.
First, I illustrate the ideal HBT radii for the blast wave parametrization
in Fig.\,\ref{idealhbt}(a)  
The model parameters are again to chosen to approximately reproduce 
the observed radii which are illustrated in Fig.\,\ref{idealhbt}(b) for comparison. 
The viscous correction to each radius is again found by substituting 
$f = f_{o} + \delta f$ into  Eq.\,\ref{fave} and expanding 
the numerator and denominator to first order in $\delta f$  
and calculating the integrals numerically.
The resulting viscous corrections are illustrated in 
Fig.\,\ref{vishbt}.
Several observations are immediate. First, as discussed 
in Sect. \ref{BjorkenSect}, the viscous corrections
in the longitudinal directions 
reduce  $\langle \tilde{z}^2 \rangle$ and $\langle \tilde{t}^2 \rangle$ 
due to the reduction of longitudinal pressure. This reduces the 
$R_{O}$ and $R_{L}$ radii.
From a phenomenological point of 
view the reduction of $R_L$ is welcome
In full ideal hydrodynamic simulations of heavy ion collisions assuming
boost invariance in the longitudinal direction \cite{KolbHBT,BassDumitru},  
$R_L$  is  approximately twice too large compared to the data.
In the blast wave model, viscous 
corrections to $R_L$ are large. This 
suggests that viscosity is responsible for the  
shortcomings in these simulations.
Comparing Fig.~\ref{idealhbt}(b) and Fig.~\ref{vishbt}(b), it seems 
that the reduction to  $R_L$   
 is too large. However, it should be 
remembered that the parameters of the blast wave model have 
been adjusted to reproduce the ideal results and 
therefore viscous corrections make the agreement with data worse. 
Further, because the correction to the longitudinal radius is large 
the calculation can not be considered reliable. For $\Gamma_s/\tau_{o}\approx 0.1$
the viscous correction to $R_L$ is approximately $30-50\%$
and the calculation is more reliable.

Viscous corrections to the transverse
variances $\langle \tilde{x}^2 \rangle$ and $\langle \tilde{y}^2 \rangle$
are small.   Consequently, 
the sideward 
radius receives only a small viscous correction. 
Viscosity introduces no significant $x-t$ correlation 
which could influence the ratio of $R_{O}$ to $R_{S}$.
In the
blast wave model the difference between  $R_{O}$ and $R_{S}$ is 
due to the contribution $\langle \tilde{t}^2 \rangle $. Viscous 
corrections to $\langle \tilde{t}^2 \rangle$ are negative and 
are essentially linearly proportional to 
this variance. 
For the particular value of $\Gamma_s/\tau_{o}=1/5$ the viscous correction
is accidentally correct and makes $R_{O}/R_{S} \approx 1$ as 
illustrated in Fig. \ref{vishbt}(b).  The agreement is accidental but
the trend is completely general. 
Viscosity reduces the $\langle \tilde{t}^2 \rangle$
and therefore tends to make $R_{O}$ equal to $R_{S}$. This 
is also welcome from a phenomenological point of view.  Full ideal hydrodynamic 
simulations (with \cite{KolbHBT,BassDumitru}  and without \cite{Hirano} the 
assumption of boost invariance) 
predict $R_{O}/R_{S} \approx 1.3$ which should be compared to 
$\sim 0.9$ observed in the RHIC data. 

In spite of these welcome corrections, including viscosity makes some aspects of the  hydrodynamic description    
of HBT radii worse.  All of the observed radii (denoted generically as $R_X$) scale quite accurately with 
$m_T=\sqrt{K_T^2 + m^2} $  as 
\begin{equation}
\label{mtscale}
            R_{X} \propto \frac{1}{\sqrt{m_T}} \; .
\end{equation}
Ideal hydrodynamics readily predicts this $\frac{1}{\sqrt{m_T}}$ 
scaling (see e.g. \cite{Sinyukov,Csorgo1,Csorgo2}). 
Indeed,  expanding Eq.~\ref{HBTideal} for the longitudinal radius 
of an ideal boost invariant expansion, we obtain the Sinyukov-Makhlin formula 
\cite{Sinyukov}  
\begin{equation}
    (R_L^2)^{(0)} = \tau_o^2 \frac{T}{m_T} \;.
\end{equation}
Viscous terms immediately break 
this $\frac{1}{\sqrt{m_T}}$ scaling. 
Expanding Eq.~\ref{HBTBj} for the longitudinal radius with viscous corrections,
we obtain
\begin{equation}
    (R_L^2)^{(0)} + \delta R_L^2 = \tau_o^2 \left(\frac{T}{m_T} - \frac{19}{16}\frac{\Gamma_s}{\tau_o} \right) \;.
\end{equation}
Viscous terms break the ideal $\frac{1}{\sqrt{m_T}}$ scaling and this
correction grows like $\frac{m_T}{T}$ relative to the ideal result. 
This deviation from  $\frac{1}{\sqrt{m_T}}$ scaling is not seen in the data. 

There remain several puzzling
aspects in the HBT measurements for which viscosity offers no 
explanation. All of the radii are the same order of 
magnitude and fall with $m_T$ as in Eq.~\ref{mtscale}. 
In particular the steep fall 
with $m_T$ in the sideward radius was difficult  to 
reproduce with the viscous blast wave model described here and 
in the ideal blast wave model \cite{Jane}.  
This behavior was predicted based upon a parametrization of ideal 
hydrodynamics \cite{Csorgo1,Csorgo2} where 
system cools rapidly during freezeout and where 
temperature and velocity gradients are 
much larger than the geometric size of the system. 
It is natural to ask whether these conditions can
be dynamically generated from some initial conditions or 
freezeout dynamics -- see \cite{Csorgo3} for efforts in 
this direction. Large velocity gradients and temperature
inhomogeneities should increase the relative importance of viscosity.
Nevertheless, the success of these models should be noted.

\section{Conclusions}

In conclusion, I have calculated the first correction 
to the thermal distribution function of an expanding gas due to 
shear viscosity. The momentum range which is accurately 
described by hydrodynamics
is directly related to the shear viscosity and 
depends upon the particular observable. I have estimated this
momentum range for single particle spectra, elliptic flow, and 
HBT radii using the boost invariant blast wave model.

For reasonable 
values of $\Gamma_{s}\equiv \frac{4}{3} \frac{\eta}{e + p}$, 
the viscous correction to the single particle 
spectrum of a blast wave model becomes of 
order one for $p_{T} \approx 1.5-2.0\,\mbox{GeV}$  as 
illustrated Fig.\,\ref{figSpectra}. 

The observed elliptic flow places a constraint on the shear viscosity. 
Indeed,  unless $\Gamma_{s}/\tau_o$ is less than $0.1$, $v_{2}$ as 
a function of $p_{T}$ falls well below the ideal curve by 
$p_{T}\approx 1.0\,\mbox{GeV}$. For the blast wave model, 
the viscous corrections to elliptic  observables become large 
$before$ the corresponding corrections  to the transverse momentum spectra. 

Shear viscosity also plays an important role in the interpretation of the 
longitudinal radius. Indeed, $R_{L}$ reflects not only the lifetime
of the system but also  the degree of thermalization in the longitudinal
direction.
$R_{L}$ involves the second moment of the thermal distribution function
in the longitudinal direction where non-equilibrium effects are the largest.
Consequently, viscous corrections to this radius 
(approximately 50\% for $\Gamma_{s}/\tau_o\approx0.2$ and 
25\% for $\Gamma_s/\tau_o\approx0.1$ .)
are  large enough  that  perhaps
$R_{L}$ should be left out of hydrodynamic fits to heavy ion data. This 
does not imply that hydrodynamics must be abandoned. On the contrary, while
thermodynamics might accurately describe $\langle p_T \rangle$, it
certainly does not accurately describe $\langle p_{T}^{100} \rangle$ unless
the viscosity is very small.  
In addition, viscous corrections to the ideal longitudinal radius 
seem to contradict measurements of $R_L$. Shear corrections 
cause the longitudinal radius to deviate from the $\frac{1}{\sqrt{m_T}}$ scaling clearly seen in the data \cite{PhenixQM2002,Jane,StarQM2002}
and expected in ideal hydrodynamics \cite{Sinyukov}.

Shear viscosity also reduces the ratio of $R_O$ to $R_{S}$ 
by decreasing the emission duration $\langle \tilde{t} \rangle$. 
Nevertheless, viscosity is not a panacea for the HBT problem.
The sideward radius falls precipitously as a function of $K_{T}$. This
precipitous fall can not be reproduced by hydrodynamics at least with 
a boost invariant expansion \cite{Csorgo}.  Viscous corrections 
to $R_{side}$  are small and make the sideward radius increase with
$K_{T}$. 

Many of the conclusions in this work about HBT radii were recently reached
``from the opposite end''
by Gyulassy and Molnar (GM) \cite{denes2}
 using kinetic theory. 
 GM, started from the Knudsen limit, increased
the transport opacity and increased the  longitudinal 
radius. 
Here, I started from the ideal hydrodynamics,  
increased the viscosity and reduced of the longitudinal 
radius.  These authors
also emphasized the importance of 
the $y-\eta_{s}$ correlation in determining $R_{L}$.  They 
also found only small viscous corrections to $R_{s}$ and experienced
similar difficulties in reproducing the steep fall in $K_T$.

Clearly performing a full viscous calculation is 
the next step towards a complete thermodynamic description  of the
heavy ion reaction.  Whether the shear viscosity can be made small 
enough ($\Gamma_s/\tau_o \lesssim 0.1$) 
in the early stages to reproduce the elliptic flow  but still 
large enough ($\Gamma_s/\tau_o \approx 0.2$) in the late stages to 
reproduce $R_{L}$ and $R_{O}/R_{S}$  remains an open 
and important dynamical question.

\noindent {\bf Ackowledgements:} I would like to thank Adrian Dumitru, Larry McLerran, Rob Pisarski, Edward Shuryak, and Raju Venugopalan  for support.  I would like to thank Denes Molnar for a careful reading of this manuscript.
This work was supported by DE-AC02-98CH10886.

\appendix
\section{The Viscous Tensor and Blast Wave Model}
\label{differentiate}

To write down  the viscous tensor 
$\left\langle \nabla_{\alpha}u_{\beta} \right\rangle$
it is most convenient to 
use Bjorken coordinates: 
$\tau=\sqrt{t^2 - z^2},\,\eta_{s}=\frac{1}{2}\log\left(\frac{t + z}{t -z}\right),\,r=\sqrt{x^2 + y^2},$ and $\phi = \mbox{atan}\left(y/x\right)$.  Note, we denote the space-time rapidity with $\eta_{s}$ and the viscous coefficient with $\eta$. 
However, we will drop the ``s'' on raised and lowered space-time indices when confusion
can not arise. In this
coordinate system the metric tensor is  
\begin{equation}
 g_{\mu\nu} = \bordermatrix{ & \tau & \eta_{s} & r & \phi \cr
  \tau & 1 & 0 & 0 & 0 \cr
  \eta_{s} & 0 & -\tau^2 & 0 & 0 \cr
  r    & 0 & 0 & -1  & 0  \cr
  \phi & 0 & 0 & 0  & -r^2 \cr}
\end{equation}
The only non-vanishing Christoffel symbols  are
$ \Gamma^{\tau}_{\eta\eta} = \tau ,
 \Gamma^{\eta}_{\tau\eta} = \frac{1}{\tau},
 \Gamma^{r}_{\phi\phi} = -r,
 \Gamma^{\phi}_{r\phi} = \frac{1}{r} $.

Without particle number conservation, the hydrodynamic fields 
are  $T(\tau,\eta_{s},r,\phi)$ and 
$u^{\mu}(r,\eta_{s}, r, \phi)$, where  $\mu=r,\tau,\eta_{s},\phi$.
The velocity field satisfies $u^{\mu} u_{\mu} = 1$ and therefore only
three components of $u^{\mu}$ need to be specified. For boost 
invariant flow $u^\eta=0$. For rotationally invariant
flow $u^{\phi}=0$. For non-rotationally  invariant 
flow we shall leave $u^{\phi}=0$ and leave the temperature
profile rotationally invariant.  
We assume boost invariance
throughout. By assumption, the particles freezeout at a proper 
time $\tau_{o}$ with a uniform distribution in the transverse plane 
and a linearly rising flow profile.  
Thus, the hydrodynamic
fields are parameterized as
\begin{subequations}
\label{flow}
\begin{eqnarray}
     T(\tau_o, \eta_{s}, r, \phi) &=&T_{o} \, \Theta (R_{o} - r)  \\
     u^{r}(\tau_o,\eta_s, r, \phi) &=& 
                        u_{o} \frac{r}{R_{o}} (1 + u_2 \cos(2\phi)) \, \Theta (R_{o}-r) \\
     u^{\phi} &=& 0 \\
     u^{\eta} &=& 0 \\
     u^{\tau} &=& \sqrt{1 + (u^{r})^2} \; .
\end{eqnarray}
\end{subequations}
For central collisions $u_2$ is zero. It is useful 
to realize that $\tau u^{\eta}$ and $r u^{\phi}$ are 
the velocities in the $\eta$ and $\phi$ directions respectively.

The viscous tensor is constructed with the 
differential operator $\nabla^{\alpha} = \Delta^{\alpha\beta}d_{\beta}$, 
where $\Delta^{\alpha\beta}$ denotes the projector, 
$g^{\alpha\beta} - u^{\alpha}u^{\beta}$, and
$d_{\beta}$ denotes the the covariant derivative, 
$d_{\beta}u^{\alpha} = 
\partial_{\beta} u^{\alpha} + \Gamma^{\alpha}_{\mu\beta} u^{\mu}$.
With these definitions the viscous tensor is given by 
$\eta\left\langle \nabla_{\alpha}u_{\beta} \right\rangle$, where  
$\left\langle \nabla_{\alpha}u_{\beta} \right\rangle \equiv 
\nabla_{\alpha}u_{\beta} + \nabla_{\beta}u_{\alpha} - 
\frac{2}{3} \Delta_{\alpha\beta}\nabla_{\gamma}u^{\gamma}$. Assuming
boost invariance, the spatial
components of the viscous tensor are given by
\begin{subequations}
\label{vistensor}
\begin{eqnarray}
r \left\langle \nabla^{r} u^{\phi} \right\rangle
 &=& -r \partial_r u^{\phi} - \frac{1}{r}\partial_{\phi} {u^{r}} -
   r u^{r} Du^{\phi} - r u^{\phi} Du^{r}
   - \frac{2}{3} r\Delta^{r\phi} \frac{1}{\sqrt{-g}} \partial_{\mu} (\sqrt{-g} u^{\mu} ) \\
r^2 \left\langle \nabla^{\phi}u^{\phi} \right\rangle
 &=& -2 \partial_\phi u^{\phi} - 2 \frac{u^{r}}{r} 
   - 2 r^2 u^{\phi} Du^{\phi} 
 -\frac{2}{3} r^2 \Delta^{\phi\phi} \frac{1}{\sqrt{-g}} \partial_{\mu} (\sqrt{-g} u^{\mu} ) \\
\left\langle \nabla^{r}u^{r} \right\rangle
 &=& - 2 \partial_r u^{r} 
   - 2 u^{r} Du^{r} 
   - \frac{2}{3} \Delta^{rr} \frac{1}{\sqrt{-g}} \partial_{\mu} (\sqrt{-g} u^{\mu} ) \\
\tau^2 \left\langle \nabla^{\eta}u^{\eta} \right\rangle
 &=& - 2 \frac{u^{\tau}}{\tau}  
   + \frac{2}{3} \frac{1}{\sqrt{-g}} \partial_{\mu} (\sqrt{-g} u^{\mu} ) \\
\left\langle \nabla^{r}u^{\eta} \right\rangle&=&
\left\langle \nabla^{\phi}u^{\eta} \right\rangle= 0  \; .
\end{eqnarray} 
\end{subequations}
Here $\sqrt{-g} = \tau r$, the expansion scalar is given by
\begin{equation}
\label{expansion-scalar}
   \frac{1}{\sqrt{-g}} \partial_{\mu} (\sqrt{-g} u^{\mu} ) = \frac{u^{\tau}}{\tau} + \frac{u^{r}}{r} + \partial_{\phi} u^{\phi} + \partial_{r} u^{r} + \partial_{\tau}u^{\tau} ,
\end{equation}
and the time derivatives in the rest frame  
$Du^{\mu}= u^{\alpha}d_{\alpha}u^{\mu}$  are given by
\begin{eqnarray}
Du^{r} &=& u^{\tau}\partial_{\tau}u^r + u^r\partial_{r}u^{r} + u^{\phi}\partial_{\phi} u^{r} - r (u^\phi)^2 \\
rDu^{\phi} &=& u^{\tau}\partial_{\tau}(ru^{\phi}) + u^r\partial_{r}(ru^{\phi}) + 
               u^{\phi}\partial_{\phi}(ru^{\phi}) + u^{\phi} u^{r} \; .
\end{eqnarray}
Once the spatial  components of the viscous stress energy tensor are known 
the temporal components are determined (numerically) from the relations, 
$\left\langle \nabla^{\alpha}u^{\beta} \right\rangle u_{\beta} = 0$.

In these equations the time derivatives,  
$\partial_{\tau} u^{\phi}, \partial_{\tau} u^{r}$, and $\partial_{\tau} u^{\tau}$ appear.
To fix the value of these time derivatives it is sufficient to 
consider the ideal equations of motion. Inclusion of viscous
terms would lead to previously neglected 
second order corrections in $\frac{\Gamma_{s}}{\tau}$.
The ideal equations of motion can be written 
\begin{eqnarray}
  De &=& - (e + p) \, \nabla_{\mu} u^{\mu} \\
  Du^{\mu} &=& + \frac{\nabla^{\mu} p}{e + p} \; .
\end{eqnarray}
With these two equations for $De$ and $Du^{r}$,
 and the flow profile given in Eqs. \ref{flow}, the
 time derivatives can be determined
\begin{subequations}
\label{tauderiv}
\begin{eqnarray}
  \partial_{\tau} u^{\phi} &=& 0 \\
  \partial_{\tau} u^{r}    &=& \frac{c_{s}^2 v}{1 - c_s^2 v^2 }\left(\frac{u^{\tau}}{\tau} + \frac{u^{r}}{r} + \partial_{r} u^{r} + v^2 \partial_{r} u^{r} \right)  - v \partial_{r} u^{r} \\
  \partial_{\tau} u^{\tau} &=& v \partial_{\tau} u^{r}  \; .
\end{eqnarray}
\end{subequations}
Here $v= u^{r}/u^{\tau}$ is the radial velocity and $c_{s}^2 = \frac{dp}{de}$ denotes the
squared speed of sound. $c_s^2$ is very close to $\frac{1}{3}$ for the pion
gas considered and is found by differentiating
the equation of state for a single component massive classical ideal gas.
See e.g. \cite{Raju} for explicit formulas for the pressure and energy 
density.
With the necessary time derivatives, the 
full viscous tensor can be found by substituting the flow profile
given in Eq. \ref{flow} into Eq. \ref{vistensor} and differentiating.  
The final formulas are lengthy and are not given. A check of the
algebra is provided by  the trace relation,  $g_{\mu\nu}T_{vis}^{\mu\nu} = 0 $.

An additional prescription for fixing the time
derivative was tried. If the particles are freezing out, then the particles 
are free streaming. Accordingly,  we have $Du^{\mu} = 0$. This  amounts 
to dropping terms proportional to $c_{s}^2$ when computing  
Eq. \ref{tauderiv}.  This
change made only a negligible change to final results. This is because the
whole effect of the time derivative is proportional to $c_{s}^2 v^{2}$ which
is rather small in practice, $c_{s}^2 v^{2} \approx \frac{1}{10}$.

To finish computing the viscous correction $p^{\mu}p^{\nu}\langle \nabla_{\mu} u_{\nu} \rangle$
we need to express $p^{\mu}$ and the integration measure $p^{\mu} d\Sigma_{\mu}$ 
in the ($\tau,\eta_{s},r,\phi$) coordinate system.  For
a particle at point  $(\tau,\eta_{s}, r,\phi)$ with  
four momentum
$p^{\mu} = (E, p^{x}, p^{y}, p^{z}) = (m_{T} \cosh y, p_{T} \cos \phi_{p},  
p_{T} \sin \phi_{p}, m_T \sinh y )$
we have
\begin{subequations}
\label{pmu}
\begin{eqnarray} 
    p^{\tau} &=& m_{T} \cosh(y-\eta_{s})  \\
    \tau p^{\eta} &=& m_{T} \sinh(y-\eta_{s}) \\
    p^{r}   &=& p_{T} \cos(\phi_{p} - \phi) \\
    r p^{\phi}   &=& p_{T} \sin(\phi_{p} - \phi) \; .
\end{eqnarray}
\end{subequations}
The oriented freezeout volume is  
$d\Sigma_{\mu}=(d\Sigma_\tau, d\Sigma_r, d\Sigma_\phi, d\Sigma_\eta) = (
\tau d\eta_s\,r dr\,d\phi,0,0,0)$ and   
the integration measure is 
\begin{eqnarray}
\label{measure}
p^{\mu} d\Sigma_{\mu} &=& m_{T} \cosh(y-\eta_{s})\,\tau d\eta_{s}\,r dr\,d\phi 
\;.
\end{eqnarray} 
With these formulas there is ample information to compute the viscous
correction  $p^{\mu}p^{\nu}\langle \nabla_{\mu} u_{\nu} \rangle$  and to
perform the necessary Cooper-Frye integrals.

\section{Viscous Corrections to a Bjorken expansion}
\label{BAppend}

In this appendix I  provide the details leading to the 
viscous corrections to the spectrum and longitudinal radius  
(Eqs.\,\ref{spectra} and \ref{HBTBj})  
for a boost invariant expansion without transverse flow.
The spectrum is given by the Cooper-Frye formula, Eq. \ref{CooperFrye}. 
First we compute the ideal spectrum. For a boost invariant expansion without 
a transverse flow $u^{\tau}=1$ and 
$u^{\eta}=u^{r}=u^\phi=0$. The thermal distribution 
for an expanding  Boltzmann gas is 
$f_{o}\left(\frac{p\cdot u}{T}\right)  = \exp\left(-\frac{m_T \cosh(y-\eta_s)} {T} \right)$.  
Then the Cooper-Frye integral gives the 
thermal spectrum from an expanding cylinder
\begin{eqnarray}
    \frac{
          d^{2}N^{(0)}
         }  { d^{2}p_{T}\,dy} &=&  
    \frac{1}{(2\pi)^3} \int p^{\mu} d\Sigma_{\mu} \, f_{o}\left(\frac{p \cdot u}{T}\right) 
\end{eqnarray}
Substituting  the integration measure $p^\mu d\Sigma_\mu$ we have
\begin{eqnarray}
\label{FOexample}
    \frac{d^{2}N^{(0)}}{d^{2}p_{T}\,dy} &=&  
    \frac{1}{(2\pi)^3} \int_{0}^{R_o} r\,dr\,\int_{0}^{2\pi} d\phi \int_{-\infty}^{\infty} \tau\, d\eta_s \,
    m_{T}\cosh(y -\eta_{s})\,  f_{o}(\frac{p\cdot u}{T})  \; .
\end{eqnarray}
Performing the integral we obtain the ideal thermal spectrum
\begin{eqnarray}
\label{dN0}
    \frac{d^{2}N^{(0)}}{d^{2}p_{T}\,dy} &=&  
    m_{T} \tau_{o} \,\frac{\pi R_o^2 }{(2\pi)^3} 
    \,2 K_1(x)  \; .
\end{eqnarray}
Here $K_{1}(x)$ is the modified Bessel function evaluated 
at $x\equiv \frac{m_T}{T}$.
Now we determine the correction spectrum. For a pure boost invariant expansion
the non-vanishing components of viscous tensor
$\left\langle  \nabla^{\mu} u^{\nu} \right\rangle$ are from Eqs. \ref{vistensor}
\begin{subequations}
\label{vistensorbj} 
\begin{eqnarray}
       \left\langle \nabla^{r} u^{r} \right\rangle &=& 
            \frac{2}{3\tau} \\
       r^2 \left\langle \nabla^{\phi} u^{\phi} \right\rangle &=& 
            \frac{2}{3\tau} \\
       \tau^2 \left\langle \nabla^{\eta} u^{\eta} \right\rangle &=& 
            -\frac{4}{3\tau} \;.
\end{eqnarray}
\end{subequations}
Thus the viscous correction $\delta f$ is
\begin{eqnarray} 
\label{deltafbj}
\delta f =
\frac{3}{8}
\frac{\Gamma_s}{T^2}\,
f_{o} \left( \frac{p\cdot u}{T} \right) 
p^{\mu}p^{\nu} \,
\left\langle \nabla_{\mu} u_{\nu} \right\rangle  
&=& 
\frac{3}{8}
\frac{\Gamma_s}{T^2}\,
 f_o \left( \frac{p \cdot u}{T} \right) 
\left( \frac{2\,p_T^2}{3\tau} - \frac{4 m_T^2 }{3\tau} \sinh^2\eta_{s} \right)
\end{eqnarray}
Note we have substituted $f_o(1+f_o)$ in Eq.\,\ref{deltaf} by $f_o$  
as required by the Boltzmann approximation.
We can then substitute $\delta f$ to determine the first viscous
correction 
\begin{eqnarray}
    \frac{
          d^{2}N^{(1)}
         }  { d^{2}p_{T}\,dy} &=&  
    \frac{1}{(2\pi)^3} 
    \int p^{\mu} d\Sigma_{\mu} \delta f \;  .
\end{eqnarray}
Substituting the integration measure 
and performing the integral over the $\eta_{s}$ as for the ideal case
we obtain
\begin{eqnarray}
\label{dN1}
    \frac{
          d^{2}N^{(1)}
         }  { d^{2}p_{T}\,dy} &=&  
    m_{T} \tau_{o} \,\frac{\pi R_o^2 }{(2\pi)^3} 
    \,2 K_1(x) \, \frac{\Gamma_s}{4\tau} 
    \left( 
     \left(\frac{p_T}{T}\right)^2  
    - \left(\frac{m_T}{T}\right)^2 \left(
    \frac{K_3(x)}{K_1(x)}  - 1\right)
    \right) \; .
\end{eqnarray}
Dividing Eq. \ref{dN1} with Eq. \ref{dN0} we obtain Eq.\,\ref{spectra} given
the text.

Next we work out the first viscous correction to the longitudinal 
HBT radius. The longitudinal radius is given by Eq.\,\ref{CFBj}. Expanding
to first order in $\delta f$ and using the relation $z=\tau_{o} \sinh \eta_s$ 
we obtain the ideal contribution
\begin{eqnarray}
    (R_{L}^2)^{(0)} (K_T) &=&  
    \frac{ 
    \int K^{\mu} d\Sigma_{\mu} \,  f_{o} (\frac{K\cdot u}{T}) \,\tau_{o}^2 \sinh^2 \eta_s 
    } { 
    \int K^{\mu} d\Sigma_{\mu} \, f_{o} (\frac{K\cdot u}{T})  \; ,
    } 
\end{eqnarray}
and the first viscous correction
\begin{eqnarray}
\label{RL1setup}
    \delta R_{L}^2 (K_T) &=& (R_L^2)^{(0)} \left(  
    \frac{  
         \frac{dN^{(1)}}{K_T dK_T}
         }{
         \frac{dN^{(0)}}{K_T dK_T}
         } \right)  + 
    \frac{ 
    \int K^{\mu} d\Sigma_{\mu} \, \delta f \tau_{o}^2 \sinh^2 \eta_{s}
    } { 
    \int K^{\mu} d\Sigma_{\mu} \, f_{o} (\frac{K\cdot u}{T}) 
    } \; .
\end{eqnarray}
For the kinematics of typical HBT measurements  at mid rapidity, we have
$K^{\mu} = (K^{\tau}, K^{r}, K^{\phi}, K^{\eta}) = (\sqrt{K_T^2 + m^2}, K_T, 0, 0)$ . 
The integration measure is 
$K^{\mu} d\Sigma_{\mu} = m_T \cosh (\eta_s)\, \tau \,d\eta_s\,rdr\, d\phi$ where 
$m_T = \sqrt{K_T^2 + m^2}$.  

First we work out the ideal radius, $(R_L^2)^{(0)}$. 
Substituting $K^{\mu}d\Sigma_{\mu}$ into the numerator and denominator 
and performing the integrals 
over the freezeout surface (as in Eq. \ref{FOexample}) we obtain 
the Herrmann-Bertsch formula \cite{Bertsch}
\begin{eqnarray}
\label{RL0}
   (R_L^2)^{(0)} = \tau_{o}^2 \frac{T}{m_T} \frac{K_2(x)}{K_1(x)} \; ,
\end{eqnarray}
where $x\equiv \sqrt{m^2 + K_T^2}/T$.  For large values of $x$, Eq.~\ref{RL0}  
reduces to the Makhlin-Sinyukov formula \cite{Sinyukov}
\begin{eqnarray}
   (R_L^2)^{(0)} = \tau_{o}^2 \frac{T}{m_T} \; .
\end{eqnarray}

A similar calculation gives the viscous 
correction. Substituting the viscous correction $\delta f$ (Eq.\,\ref{deltafbj}) into
Eq.\,\ref{RL1setup}, using the previous results for the 
spectrum (Eqs. \ref{dN0}, \ref{dN1}) and ideal radius (Eq. \ref{RL0}), 
and performing the  $\eta_s$ integrals, we obtain Eq.\, \ref{HBTBj} 
quoted in the text
\begin{eqnarray}
   \frac{\delta R_L^2} {(R_L^2)^{(0)} }  &=&
-\frac{\Gamma_s}{\tau} 
\left(
       \frac{6}{4} 
      \frac{ 
        x K_3(x)
      }{ 
         K_2(x)
      }  -
      x^2 \frac{1}{8} 
      \left( \frac{
                    K_3(x)
                  } {
                    K_2(x)
                  } - 1 
      \right) 
\right)\; .
\end{eqnarray}
\end{document}